\documentclass[twocolumn,twocolappendix]{aastex701}

\usepackage{array}
\usepackage{graphicx, url}
\usepackage{hyperref, amsmath}
\usepackage{multirow}
\usepackage{natbib}
\usepackage{xcolor}
\usepackage{soul}

\definecolor{dark green}{rgb}{0.00, 0.39, 0.00}

\def\D{\,\textrm{d}}
\def\bh{\,\textrm{BH}}

\newcommand{\eref}[1]{Eq.~(\ref{#1})}
\newcommand{\esref}[2]{Eqs.~(\ref{#1}) and~(\ref{#2})}
\newcommand{\fref}[1]{Fig.~\ref{#1}}

\newcommand{\sref}[1]{Sec.~\ref{#1}}
\newcommand{\tref}[1]{Table~\ref{#1}}

\shorttitle{Extreme Values of Black Hole to Stellar Mass Ratio}
\shortauthors{C.Heather et al}

\begin{document}


\title{Extreme Values of Black Hole to Stellar Mass Ratio for High-Redshift Galaxies}

\author{Cameron Heather}
\affiliation{Warwick Mathematics Institute, University of Warwick, Zeeman Building, Coventry CV4 7AL, UK}
\email[show]{cameron.heather@warwick.ac.uk}  

\author{Teeraparb Chantavat} 
\affiliation{Institute for Fundamental Study, Naresuan University, Phitsanulok, 65000, Thailand}
\email{teeraparbc@nu.ac.th}

\author{Siri Chongchitnan}
\affiliation{Warwick Mathematics Institute, University of Warwick, Zeeman Building, Coventry CV4 7AL, UK}
\email{siri.chongchitnan@warwick.ac.uk}

\author{Josesph Silk}
\affiliation{Institut d'Astrophysique de Paris, 98 bis Boulevard Arago, 75014, Paris, France}
\affiliation{William H. Miller III Department of Physics and Astronomy,The Johns Hopkins University, Baltimore, MD 21218, USA}
\email{silk@iap.fr}

\begin{abstract}
With recent data from the \emph{James Webb Space Telescope} (JWST), it is possible to calculate the mass of the supermassive black holes at the center of galaxies, and the stellar mass of the host galaxies at redshift $z \gtrsim 5$. In this work, we apply extreme-value statistics to calculate the distributions of extreme black-hole and stellar masses for galaxies in the redshift range $3 \lesssim z \lesssim 8$. We show that under certain assumptions about the stellar and black-hole mass functions, a high ratio of $M_\text{BH}/M_*\sim0.3-0.5$ can be obtained without invoking additional black-hole growth physics. Nevertheless, surveying a range of extreme-value methodologies, we find predictions of the extreme ratio  $M_\text{BH}/M_*$ to still be in slight tension with the high values observed by JWST.
\end{abstract}


\section{Introduction}

The \textit{James Webb Space Telescope} (JWST) has now revealed hundreds of galaxies at redshift $z\gtrsim5$, shedding new light on galaxy evolution in the Universe less than a billion years after the Big Bang (e.g., \cite{Finkelstein_ea2023, Naidu_ea2022}). Using JWST NIRSpec and NIRCam, supermassive black holes (SMBHs) at the center of these high-redshift galaxies can be probed, and their masses inferred from broad emission lines (e.g., \cite{Harikane_ea2023, Maiolino_ea2024}). In addition, the stellar mass in these galaxies can be estimated by spectral energy distribution (SED) fitting of host galaxies, though with significant uncertainties due to dust and nebular emission.

These observations revealed that some high-redshift SMBHs are disproportionately massive with respect to their host galaxies. Recent JWST observations found $M_{\rm BH}/M_*$ as large as $10\%$ for galaxies at $z\gtrsim5$  \citep{kokorev, matthee, yue}.   This is in contrast to the much lower, previously accepted pre-JWST values of  $(M_{\rm BH}/M_*)\sim10^{-3}$ \citep{halring, ferrarese, reines}. Several ultra-high redshifts AGN ($z>10$) suggest that galaxies could even form with over-massive SMBH with $(M_{\bh}/M_*)\sim1$ \citep{bogdan, pacucci}

These high ratios challenge traditional models of how SMBHs and their host galaxies co-evolve, and raise questions about the physics of SMBH growth in the early Universe. A host of exotic physics has been proposed, such as super-Eddington accretion \citep{Pezzulli}, top-heavy initial mass functions \citep{haghi}, and seeding by extreme Pop III stars \citep{chantavat}. Nevertheless, a consistent theoretical framework has yet to emerge. 

In this work, we investigate whether the high values of the observed ratio $M_{\bh}/M_*$ are consistent with predictions from \textit{extreme-value statistics} (EVS). 

We present two EVS pipelines, yielding different predictions for extreme $M_{\bh}/M_*$ ratios. In the first pipeline, the EVS framework is applied to obtain the extreme values of galaxy stellar masses. The extreme black-hole masses are then estimated by virtue of previously established linear scalings between $\log M_\text{BH}$ and $\log M_*$. The result of this pipeline  (\sref{sec:gauss_evs}) show that the extreme values predictions for $ M_\text{BH}$ and $M_*$ are consistent with JWST observations, although the EVS prediction for the ratio remains  low.

In the second pipeline, the EVS framework is applied to both the stellar and black-hole mass distributions separately. We then calculate the extreme ratio assuming that extreme black holes are correlated with extreme stellar masses.  The results of this pipeline (\sref{sec:bh_ratio}) show that extreme ratios as high as $0.3-0.5$ at redshift $z\sim4-8$ can be obtained without invoking exotic growth physics.

Our conclusions, a discussion of both pipelines, and a comparison with JWST data are presented in \sref{sec:conclusion}.


Throughout this paper, we will assume a flat $\Lambda \rm{CDM}$ cosmology with $H_0 =
70 \rm{km}\ s^{-1} Mpc^{-1}$, $\Omega_{\rm m} = 0.32$ and $\Omega_{\Lambda} = 0.68$.


\section{Abundances of stars and black holes at high redshifts}

A key ingredient in the EVS formalism is the number count of rare objects. In this work, we need to quantify the stellar  and black-hole masses of galaxies at high redshift.  We obtain these abundances from the galaxy stellar mass functions given in \cite{shuntov25, Navarro-Carrera24}, and the black hole mass functions in \cite{roberts26, Taylor2024}. We focus on the redshift range $3.5 < z < 8.5$, where observations from JWST have emerged in the past few years.


\subsection{Stellar Mass Functions}\label{sec:SMF}

We now describe two galaxy stellar mass functions.
Firstly, \cite{shuntov25} analysed the wide-area COSMOS-Web surveys containing $\sim7\times10^5$ galaxies up to $z\sim12$ and obtained stellar mass $M_*$ via SED fitting.  The stellar mass function obtained was found to be well approximated by a concatenation of the Schechter function \citep{Schechter1976} at lower redshift $3.5 < z < 5.5$:
\begin{multline}\label{eqn:schechter}
    \Phi\ {\rm d}(\log M_*) = \ln(10)\ \varphi^*\exp\left[-10^{\log M_*-\log M_c}\right]\times  \\
    \left(10^{\log M_*-\log M_c}\right)^{\alpha+1}  {\rm d}(\log M_*),
\end{multline}
 and a Double Power Law at $z\geq5.5$:
\begin{multline}\label{eqn:DPL}
    \Phi\ {\rm d}(\log M_*) = \ln(10)\exp\left[-10^{\log M_*-\log M_c}\right]\times \\
    \left[\varphi_1^*\left(10^{\log M_*-\log M_c}\right)^{\alpha_1+1} + \varphi_2^*\left(10^{\log M_*-\log M_c}\right)^{\alpha_2+1}\right]\\ {\rm d}(\log M_*).
\end{multline}
We use the values of the parameters $\alpha, \alpha_1, \alpha_2, M_c$ given in Appendix G of \cite{shuntov25}. \fref{fig:SMF_plots} (top panel) shows the plot of the Shuntov stellar mass function.

Secondly, \cite{Navarro-Carrera24} analysed NIRCam images of $3300$ galaxies in the redshift range $3.5<z<5.5$ in a much narrower area, supplemented at the high mass end by the CANDELS catalogue \citep{Grogin_2011}. A stellar mass function of the Schechter form was then obtained as a good fit in \cite{Navarro-Carrera24}. This mass function shows weak redshift dependence and a non-monotonic evolution over the redshift bins $[3.5,4.5], [4.5,5.5], [5.5,6.5], [6.5,7.5], [7.5,8.5]$. We plot this mass function in the lower panel of \fref{fig:SMF_plots}.


\begin{figure}[htb!]
    \centering
    \includegraphics[width = \linewidth]{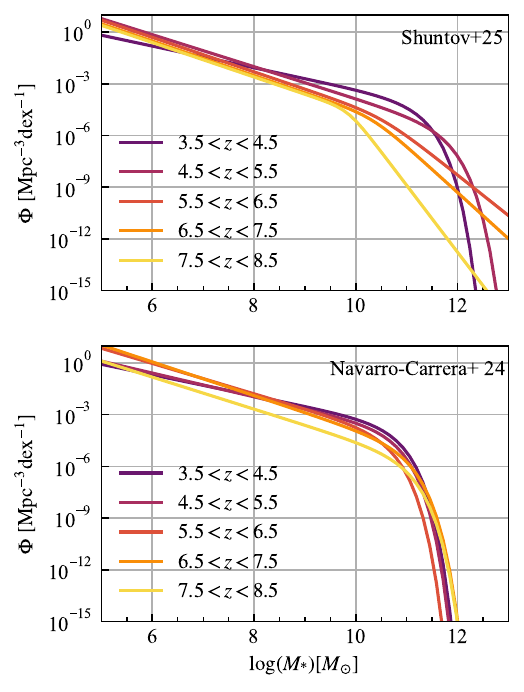}
    \caption{The stellar mass functions used in this work are from \cite{shuntov25} (top) and \cite{Navarro-Carrera24} (bottom), plotted here over the redshift $z = 3.5 - 8.5$. The Shuntov mass function is a combination of the Schechter and Double Power Law functions, whilst the Navarro-Carrera mass function takes the Schechter form. }
    \label{fig:SMF_plots}
\end{figure}

\subsection{Black Hole Mass Function}\label{sec:BH_funcs} 

In this work, the black-hole mass function is obtained in two ways. 
Firstly, following the method outlined in \cite{roberts26}, the black-hole mass function can be obtained via a Gaussian convolution of the stellar mass function:
\begin{equation}
    \Phi(M_{\bh}) = \int^{\infty}_{M_{*,\text{min}}} \Phi(M_*){\rm P}(M_{\bh}|M_*)\,\rm{d} log(M_*),
\label{convo}
\end{equation}
where  ${\rm P}(M_{\bh}|M_*)$ is a Gaussian distribution centered at the mean of the $M_{\bh}$-$M_*$ relation of the form 
\begin{equation}\label{eqn:lin_param}
    \log\left(\frac{M_{\bh}}{\rm M_{\odot}}\right) = \alpha + \beta\log\left(\frac{M_*}{\rm M_{\odot}}\right)+\epsilon.
\end{equation}
The parameters $\alpha, \beta$ and the scatter $\epsilon$ in this log-linear scaling have previously been calculated in the pioneering work of \cite{reines}, and updated with JWST observations in  \cite{Li25} and \cite{pacucci23}. Whilst the relations obtained by Li and Pacucci were fitted at $z\in[4,7]$, Reines and Volonteri obtained the scaling relation at low redshift ($z\sim0.5$). Extrapolating the latter to $z\sim4-8$ has been shown to severely underestimate black hole masses at high redshifts, and so we will use discuss the mass functions of Li and Pacucci in this work. We will return to this linear scaling in Section \ref{sec:gauss_evs}.

Alternatively, in \cite{Taylor2024}, a black-hole mass function in the redshift range $3.5<z<6$ was obtained from analysing a NIRSpec sample of broad-line AGNs, using broad H$\alpha$ emission to estimate black-hole masses. The authors found a mass function of the Schechter form that does not evolve significantly with redshift. In this work, we further extrapolate their black hole mass function to $z\sim8$, effectively assuming that the most massive SMBHs were already in place by $z\sim 8$. (We will revisit the Taylor mass function in \sref{sec:bh_ratio}.) 

\begin{figure}[htb!]
    \centering
    \includegraphics[width = \linewidth]{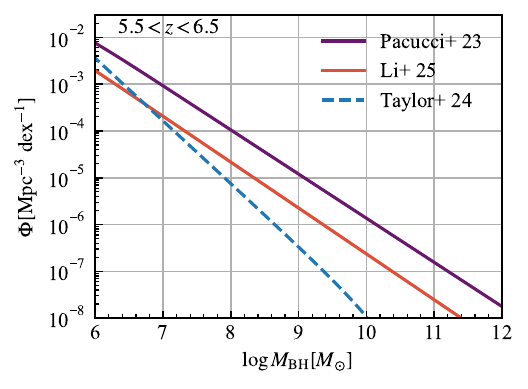}
    \caption{The black-hole mass functions used in this work, shown here for the redshift bin $5.5<z<6.5$. The solid lines are mass functions derived from convolving the Shuntov stellar mass function with a pdf with two different Gaussian profiles, namely, \cite{pacucci23} and  \cite{Li25}. The dashed line shows the black-hole mass function obtained by \cite{Taylor2024} using JWST NIRSpec data. }
    \label{fig:BH_Plots}
\end{figure}

For comparison, we plot the three black-hole mass functions used in this work in \fref{fig:BH_Plots} in the redshift bin $5.5 < z < 6.5$.

\section{Extreme-value modelling}\label{sec:EVS}

We now describe the EVS framework which gives the distribution of rare objects in a given survey volume. We will apply the EVS framework to find the distributions of  black-hole and stellar masses in high-redshift galaxies. This EVS framework has previously been used to calculate extreme halo and stellar masses of JWST galaxies \citep{Lovell2023}, the most massive Pop III stars \citep{chantavat}, the most massive primordial black holes \citep{Chongchitan2021a, Chongchitnan2021b}, and, more recently, the brightest galaxies and AGNs at high redshifts \citep{Heather2024, Heather2025}.



We will be working with the generalised extreme value (GEV) approach, also known as the block maxima method \citep{Gumbel}. In this method, the observational volume is divided into $N$ distinct blocks. The distribution of the most massive object in each block is obtained using a probability density function outlined below. 



Firstly, we calculate the number count of objects with mass greater than $M$ in a given redshift bin by integrating the mass function:
\begin{equation}\label{eqn:ncount}
    n(>M) = \int^{z_{\rm max}}_{z_{\rm min}} \int^{\infty}_{M} \frac{\D n}{\D M'}\ \D M'\ \D z,
\end{equation}
where $M$ is either black hole mass, $M_{\bh}$, or stellar mass, $M_*$ associated with a 
galaxy. 

Next, we calculate $P_0$, the probability that no galaxies in a given volume $V$ will have black hole or stellar mass exceeding $M$. We assume the observed sky fraction $f_{\rm sky}= 1$, to obtain conservative predictions on the extreme masses. The probability distribution $P_0(M)$ can be modelled as a Poisson distribution with the following cumulative distribution function (cdf),

\begin{equation}\label{eqn:P0}
    P_0(M) = \exp(-n(>M)V).
\end{equation}

In the GEV formalism, Equation (\ref{eqn:P0}) approaches the following distribution in the large-$N$ limit 
\begin{equation}\label{eqn:GEV}
    G(M) =  \begin{cases} 
    \exp \left( -(1+\gamma y)^{-1/\gamma}\right) & (\gamma \neq 0),\\
    \exp \left(-e ^{-y}\right) & (\gamma = 0).
    \end{cases}
\end{equation}
where $y = (M - \alpha)/\beta$.
The parameters $\alpha, \beta$ describe the location of the peak and the scale of the pdf, while the sign of $\gamma$ determines which GEV distribution the pdf corresponds to.  A mathematical result known as the Fisher-Tippett theorem states that in the large-sample limit, the distribution \ref{eqn:P0} converges to one of three distributions: Gumbel, Fr\'echet or Weibull distributions, corresponding to $\gamma = 0 ,\ \gamma >0,\ \gamma< 0$ respectively (see \cite{coles} for a review).

We can express the parameters $\alpha, \beta, \gamma$ in terms of $M$ and $V$ by Taylor-expanding \esref{eqn:P0}{eqn:GEV} around the peak of the pdf (at $M=M_{peak}$) to cubic order. We obtain:
\begin{align*}
    \gamma &= n(>M_{\rm peak})V - 1 \\
    \beta &= \frac{(1+\gamma)^{1+\gamma}}{\frac{\D n}{\D M}\big\rvert_{M_{\rm peak}}M_{\rm peak}V} \\
    \alpha &= \begin{cases}
    M_{\rm peak} - \frac{\beta}{\gamma}\left((1+\gamma)^{-\gamma} - 1 \right)& (\gamma \neq 0),\\
    M_{\rm peak}
    & (\gamma= 0).
    \end{cases}
\end{align*}


The main advantage of the GEV distribution is that it is a smooth 3-parameter $(\alpha, \beta, \gamma)$ description of the Poisson model, giving us a convenient way to compare extreme values in different physical models or at different redshifts. Extreme-value distributions can also be obtained by interpolating or extrapolation $(\alpha, \beta, \gamma)$ without having to integrate the mass function again. 

In addition, the sign of $\gamma$ tells us whether the effective tail is heavy, power-law like ($\gamma>0$), exponential ($\gamma=0$), or bounded above ($\gamma<0$). We expect similar physical processes to converge to the same class of GEV distribution.

\begin{figure}[htb!]
    \centering
    \includegraphics[width = \linewidth]{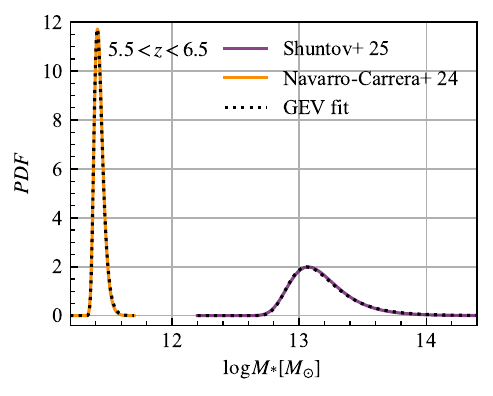}
    \caption{Extreme-value distribution for the stellar mass $M_*$ in the redshift bin $5.5<z<6.5$. These pdfs are obtained using the stellar mass functions of \cite{shuntov25} (right, broader pdf) and \cite{Navarro-Carrera24}. The solid lines are derived from the Poisson distribution \ref{eqn:P0}. These are accurately approximated by the Gumbel distributions, shown in dashed lines.
    }\label{fig:SMF_pdf}
\end{figure}

Fig. \ref{fig:SMF_pdf} shows the extreme-value distribution of stellar mass using the mass functions described in section \ref{sec:SMF} in the redshift bin $z\in[5.5,6.5]$. The Poisson-derived distributions (solid lines) are accurately fitted by the GEV fits (dashed lines). For a wide range of redshift $z\in[4,8]$, we found $\gamma$ to be negligibly small, meaning that the extreme-value distributions for $M_*$ are well approximated by the Gumbel distribution. 

The Shuntov stellar mass function was based on a wide COSMOS survey which found many more massive galaxies in a wider range of redshifts than the data used in Navarro-Carrera. The latter was based on a narrow pencil-beam survey $\sim100$ times smaller than the COSMOS survey. This explains why the Shuntov extreme pdf in fig. \ref{fig:SMF_pdf} is broader and spans bigger stellar masses than the Navarro-Carrera extreme pdf.



Similarly, these distributions were obtained in the five redshift bins spanning $3.5<z<8.5$. The profiles of the EVS distributions are shown in \fref{fig:SMF_pdf}. The 95th and 99th percentiles of the distributions are shaded accordingly. 

The Shuntov stellar mass function gives rise to more redshift fluctuations in extreme stellar mass compared to that from the Navarro-Carrera mass function. The latter shows an almost constant extreme stellar mass of around $\sim10^{11.6} M_\odot$ across the entire redshift range. This non-monotonicity stems from that in both the stellar mass functions. In particular, the characteristic mass for both \cite{shuntov25, Navarro-Carrera24} in their fitted function evolves non-monotonically with redshift. We see a change in the extreme distribution for \cite{shuntov25} quite clearly for $ z > 5.5$, which is where the change from  Schechter to double power law happens. 


\begin{figure}[htb!]
    \centering
    \includegraphics[width = \linewidth]{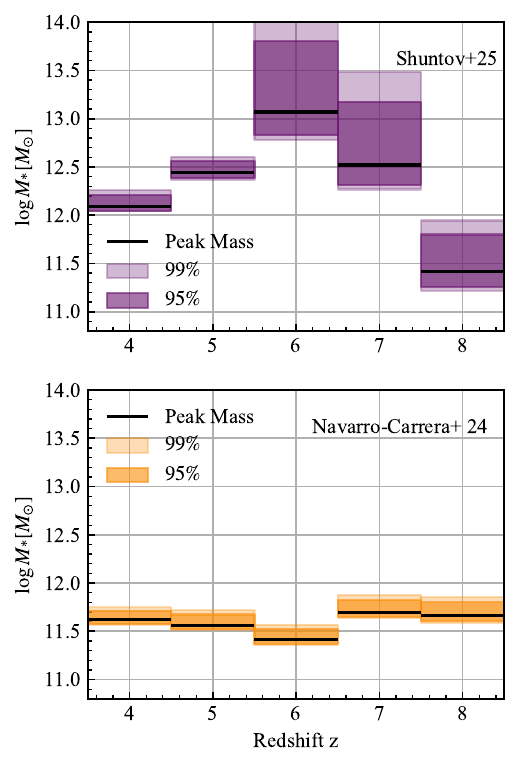}
    \caption{Extreme values of galaxy stellar masses derived from the stellar mass function of  \cite{shuntov25} (top) and  \cite{Navarro-Carrera24} (bottom). In each bin, we display the peak of the pdf (solid line), the 95th and the 99th percentiles.
    }\label{fig:SMF_EVS}
\end{figure}

\medskip

In the next two sections, we survey two EVS pipelines that can be used to obtain the extreme distributions of $M_\text{BH}$, $M_*$ and their ratio.

\section{Pipeline I: Extremes of the ratio $M_{\bh}$/$M_{*}$ via observed  correlation}\label{sec:gauss_evs}

In this pipeline, we start with the distribution of extreme stellar mass calculated in \sref{sec:EVS} using both the \cite{shuntov25, Navarro-Carrera24} stellar mass functions, plotted in \fref{fig:SMF_EVS}. We then obtain the corresponding extreme BH mass by applying the log-linear correlation (Eq. \ref{eqn:lin_param}), derived from the Gaussian convolution as explained in \sref{sec:BH_funcs}. This gives the extreme distributions for the black hole mass.  

To calculate the ratio $M_{\bh}/M_*$, we take $10^4$ samples from each of the extreme distributions of stellar and black hole mass using the inverse GEV distribution obtained from inverting \eref{eqn:GEV}. From these, we compute the median value of the ratio assuming the \cite{shuntov25} or \cite{Navarro-Carrera24} mass function. 

\begin{figure}[htb!]
    \centering
    \includegraphics[width = \linewidth]{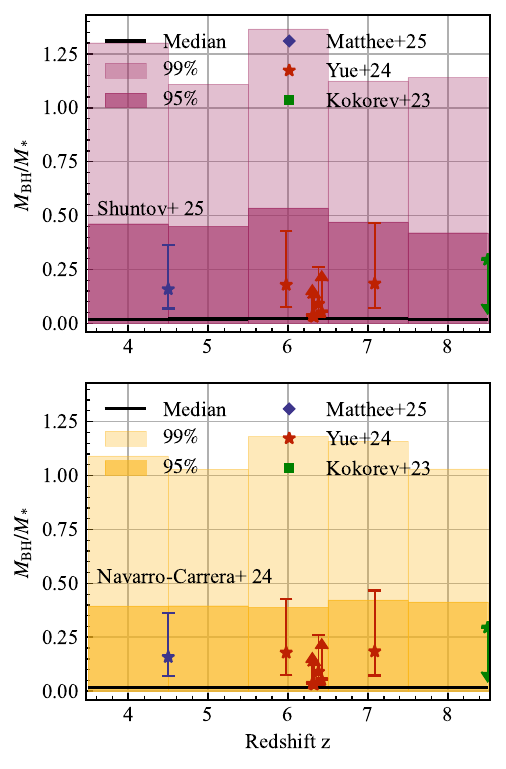}
    \caption{Plots of the EVS prediction for the extreme $M_{\bh}-M_*$ ratio for redshift bins in the range $z = 3.5 - 8.5$. The values for $M_{\bh}$ and $M_*$ are calculated by applying the linear parametrizations of the $M_{\bh} - M_*$ ratio to the extreme-value modelling of the stellar mass function from \cite{shuntov25} (top) and \cite{Navarro-Carrera24} (bottom), with parameter values from \cite{pacucci23}. The solid black line corresponds to the median, whilst the darker/lighter shaded area corresponds to the $95^{th}$/$99^{th}$ percentile. The data points are from \cite{matthee, yue, kokorev}.
    }
    \label{fig:pip1_evs}
\end{figure}

We plot the EVS prediction of the ratio in \fref{fig:pip1_evs}, including the $99^{\rm th}$ and $95^{\rm th}$ percentiles as shaded confidence intervals. We also include the observed ratios from \cite{matthee, yue, kokorev}. In \fref{fig:pip1_evs}, we show the distributions using the log-linear correlation with coeffcients given in  \cite{pacucci23}. The authors studied JWST AGNs at $z=4-7$, finding that the black holes at these redshifts are 10 - 100 times overmassive compared to low-redshift galaxies.

The EVS preditions are broad, with the $95^{\rm th}$ percentile extending to $M_{\bh}/M_*\sim0.4-0.5$, consistent with JWST data. The broad confidence region is due to a large scatter of $~0.7$ dex in the Pacucci scaling relation.

We have also performed the same EVS calculation  using the  parametrization given in  \cite{Li25}. The authors analysed JWST AGNs at $z>6$ and  concluded that selection effects, measurement uncertainties, and intrinsic scatter can explain the observed black hole detection, without requiring the population to be intrinsically overmassive compared to local relations (such as that in \cite{reines}).

The interpretation of \cite{Li25} naturally leads to an even broader (and hence less predictive) EVS distribution for the ratio $M_{\bh}/M_*$. We show the EVS plots and data for \cite{Li25} in Appendix  \ref{appendix}.


\begin{figure}[htb!]
    \centering
    \includegraphics[width = \linewidth]{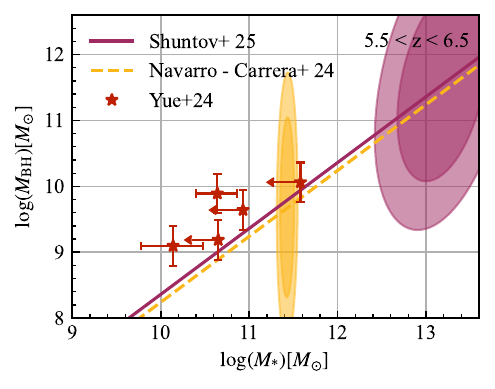}
    \caption{The ellipses show EVS predictions in $5.5\lesssim z \lesssim 6.5$ for extreme black hole mass (vertical axis) and extreme stellar mass (horizontal axis) using Pipeline I (\S\ref{sec:gauss_evs}). The pink/larger region assumes the \cite{shuntov25} mass function whilst orange/thinner region assumes the \cite{Navarro-Carrera24} mass function. The \cite{pacucci23} scaling for the $M_{\rm BH}$-$M_*$ relation is assumed.  The values for the extreme ratio (represented by the gradient of the lines) are provided in \tref{table_pip1}. We include the data points from \cite{yue}, which lie to the left and below the Shuntov ellipse, hence showing consistency between this EVS formalism and observation. There is some tension between the Navarro-Carrera ellipse and observed extreme $M_*$ in this redshift bin.
    }\label{fig:pip1_ratio}
\end{figure}

It is useful to see the predictions for extreme values of $M_{\rm BH}$ and $M_*$ separately. In \fref{fig:pip1_ratio}, we present these extremes as confidence ellipses (where $~95\%$ and $99\%$ of the samples of the extreme distributions lie), focussing on the redshift bin $5.5 < z < 6.5$.

We see that despite the \cite{shuntov25} and \cite{pacucci23} prescriptions having a similar median ratio (represented by the gradient of each straight line in this figure), the two mass functions predict very different extreme masses. In particular, \cite{shuntov25} predicts a much larger range for both the stellar and black hole masses we expect to see. While the \cite{Navarro-Carrera24} mass function gives the median ratio that is in agreement with the data, it under-predicts the values for the stellar and black hole mass themselves. To be consistent with the EVS prediction, the observed data should lie below and to the left of the confidence intervals. We see that the Navarro-Carrera prescription does not produce observationally consistent extreme-value predictions due to the mass function being calibrated to a small sample of mostly lower mass galaxies. 



\begin{deluxetable}{cccccc}[htb!]
\tablehead{
\colhead{SMF in Pipeline I} & \colhead{Redshift} & \colhead{$M_{\bh}/M_*$} & }
\startdata
\cite{shuntov25}&$[3.5,4.5]$&$0.020^{+0.442}_{-0.019}$ \\
&$[4.5,5.5]$&$0.020^{+0.429}_{-0.019}$\\
&$[5.5,6.5]$&$0.023^{+0.512}_{-0.022}$\\
&$[6.5,7.5]$&$0.021^{+0.450}_{-0.020}$\\
&$[7.5,8.5]$&$0.019^{+0.402}_{-0.018}$\\ \hline
\cite{Navarro-Carrera24}&$[3.5,4.5]$&$0.018^{+0.378}_{-0.017}$\\
&$[4.5,5.5]$&$0.018^{+0.379}_{-0.017}$\\
&$[5.5,6.5]$&$0.017^{+0.372}_{-0.017}$\\
&$[6.5,7.5]$&$0.019^{+0.405}_{-0.018}$\\
&$[7.5,8.5]$&$0.019^{+0.396}_{-0.018}$\\
\enddata
    \caption{The values of the ratio $M_{\text{BH}}/M_*$ calculated in \sref{sec:gauss_evs} in  $3.5<z<8.5$. Pipeline I uses the log-linear relation (Eq. \ref{eqn:lin_param}) with  coefficients given in \cite{pacucci23}. The stellar mass functions used are from  \cite{shuntov25} and \cite{Navarro-Carrera24}. We include the median values and the $95\%$ confidence intervals for the ratio $M_{\text{BH}}/M_*$. This data corresponds to the dark shaded regions in \fref{fig:pip1_evs}, and the gradients of the lines in \fref{fig:pip1_ratio}.}\label{table_pip1}
\end{deluxetable}

\section{Pipeline II: Ratio of extremes}\label{sec:bh_ratio}

In this alternative approach, we apply EVS to find the extreme stellar and black hole masses separately, using the corresponding mass function in the literature. We then calculate the distribution of the ratio \begin{equation}
    \mu=M_{\text{BH,max}}/M_{*,\text{max}}
\end{equation}
  by independently sampling the marginal extreme-value distributions for the black-hole and stellar masses. The result is not necessarily the distribution of extreme values of the ratio $M_{\bh}/M_*$, but rather a diagnostic calculation comparing the tails of the extreme distributions, in the absence of the joint probability distribution $P(M_\text{BH}, M_{*}, z)$, which is largely unknown.

For the stellar mass,  we already have from  \S\ref{sec:EVS} the distribution for $M_{*,\text{max}}$ assuming the \cite{shuntov25} or the \cite{Navarro-Carrera24} stellar mass function. Following the same formalism, we calculate the distribution for $M_{\text{BH,max}}$ using black hole mass function from  \cite{Taylor2024}, whose validity we briefly discuss below. 

\cite{Taylor2024} derived a broad-line AGN black-hole mass function from 62 JWST/NIRSpec AGN at $3.5<z<6$ and found a Schechter-like fit with little apparent redshift evolution. Whist this mass function is used to estimate black hole abundance in this Section, its high-mass tail should be interpreted with caution. A high abundance of massive black holes at these redshifts implies rapid early assembly and, via Soltan-type bookkeeping, requires substantial prior accretion luminosity \citep{jahnke}. Also, the Taylor mass function is based on single-epoch virial masses calibrated at low redshifts \citep{reines2}, leading some to speculate that the high mass end of the Taylor mass function could be biased upwards  (\textit{e.g.} see  \cite{rusakov} and counterarguments in \cite{juodzbalis}).

\begin{figure}[htb!]
    \centering
    \includegraphics[width = \linewidth]{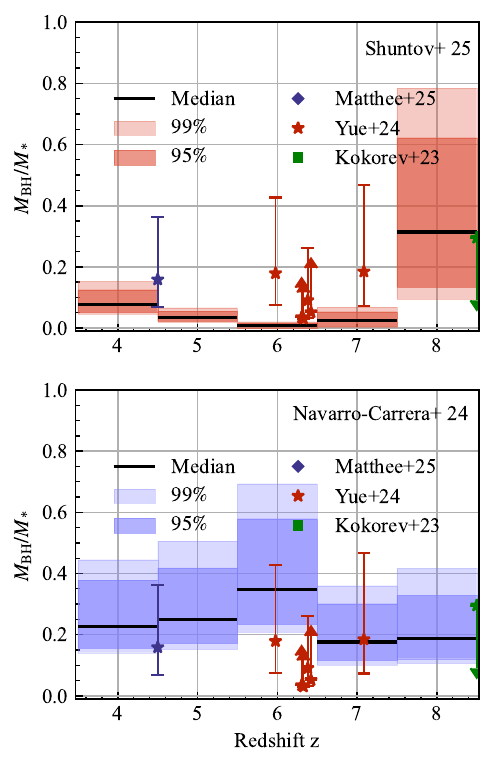}
    \caption{Plots of the EVS prediction for the ratio of extremes, i.e. $\mu=M_{\text{BH,max}}/M_{*,\text{max}}$ in the redshift range $z = 3.5 - 8.5$. The extreme values for $M_*$ are calculated using the mass functions from \cite{shuntov25} (top) and \cite{Navarro-Carrera24} (bottom) respectively. The solid black line corresponds to the median, whilst the darker/lighter shaded area corresponds to the $95^{th}$/$99^{th}$ percentile. 
    }\label{fig:pip2_evs}
\end{figure}

Having obtained the distribution for  $M_{*,\text{max}}$ and $M_{\text{BH,max}}$, we sample from each of these extreme distributions using the inverse GEV distribution, giving  $10^4$ samples of the ratio $\mu$. The median value, $99\%$ and $95\%$ confidence intervals of $\mu$ are plotted in \fref{fig:pip2_evs}, for both the \cite{shuntov25} and \cite{Navarro-Carrera24} stellar mass functions, along with observed ratios. The numerical values for these results are provided in \tref{table_pip2}.

We find the observations from \cite{matthee, kokorev} to be consistent with both prescriptions, but the observed ratio from \cite{yue} at $z=5-7$ are too high to be consistent with the prediction from \cite{shuntov25}. For example, the latter predicts $\mu\approx 0.007$, much smaller than the observed ratios, in contrast with  $\mu\approx 0.347$ in the \cite{Navarro-Carrera24} prescription.

\begin{figure}[htb!]
    \centering
    \includegraphics[width = \linewidth]{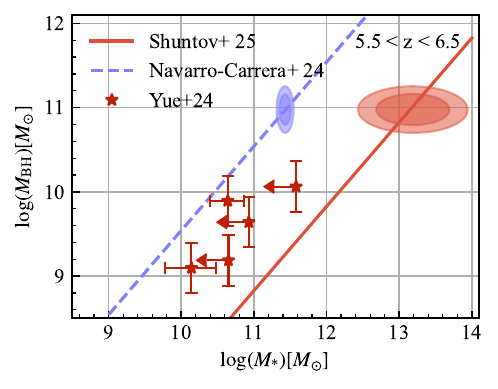}
    \caption{The ellipses show the EVS predictions in $5.5\lesssim z \lesssim 6.5$ using Pipeline II (\S\ref{sec:bh_ratio}).  The red/larger ellipse assumes the \cite{shuntov25} stellar mass function whilst the blue/smaller ellipse assume the \cite{Navarro-Carrera24} mass function. Only the latter is consistent with data. The extreme black-hole masses were calculated using the \cite{Taylor2024} mass function. The gradient of each line through the centroid corrresponds to the ratio $\mu$. }
    \label{fig:pip2_ratio}
\end{figure}

Finally, we display both extreme ratios as confidence ellipses in \fref{fig:pip2_ratio} for the single redshift bin $5.5 < z < 6.5$.  In this view, the extreme $M_*$ and $M_\text{BH}$ using both the \cite{shuntov25} and \cite{Navarro-Carrera24} prescriptions  are consistent with data. However, the ratio $\mu$ is consistent only with the \cite{Navarro-Carrera24} prescription. 

\begin{deluxetable}{ccc}[htb!]
\tablehead{
\colhead{SMF in Pipeline II} & \colhead{Redshift} & \colhead{$\mu$}  }
\startdata
\cite{shuntov25}&$[3.5, 4.5]$&$0.075^{+0.050}_{-0.024}$\\
&$[4.5, 5.5]$&$0.033^{+0.022}_{-0.011}$\\
&$[5.5, 6.5]$&$0.007^{+0.010}_{-0.005}$\\
&$[6.5, 7.5]$&$0.024^{+0.030}_{-0.018}$\\
&$[7.5, 8.5]$&$0.314^{+0.308}_{-0.179}$\\ \hline
\cite{Navarro-Carrera24}&$[3.5, 4.5]$&$0.225^{+0.153}_{-0.069}$\\
&$[4.5, 5.5]$&$0.250^{+0.168}_{-0.079}$\\
&$[5.5, 6.5]$&$0.347^{+0.232}_{-0.114}$\\
&$[6.5, 7.5]$&$0.176^{+0.126}_{-0.060}$\\
&$[7.5, 8.5]$&$0.187^{+0.141}_{-0.065}$\\
\enddata
    \caption{The values of the ratio of extremes $\mu=M_{\text{BH,max}}/M_{*,\text{max}}$ as shown in  \fref{fig:pip2_evs}. }\label{table_pip2}
\end{deluxetable}


\section{Conclusion and Discussion}\label{sec:conclusion}

\subsection{Summary of main results}

\begin{table*}[htb!]
\centering
\begin{tabular}{c||c|c}
& \multicolumn{2}{c}{\textbf{Stellar mass function}} \\ 
\textbf{Pipeline} & \cite{shuntov25} & \cite{Navarro-Carrera24} \\ \hline
I (\sref{sec:gauss_evs}) & Broad EVS predictions, & Broad EVS predictions,\\
(Log-linear correlation)& consistent with data. & some tension with data. \\
&  \fref{fig:pip1_evs} (top) & \fref{fig:pip1_evs} (bottom), \fref{fig:pip1_ratio}\\ \hline
II (\sref{sec:bh_ratio}) & Ratio too small in $z=$5-7, & Consistent with data. \\
(Ratio $\mu$)& inconsistent with data. & \fref{fig:pip2_evs} (bottom), \fref{fig:pip2_ratio}\\
&  \fref{fig:pip2_evs} (top) &  \\
\end{tabular}
    \caption{Summary of key results in this work. }\label{tab:summary}
\end{table*}

We used extreme-value statistics to predict extreme-mass black hole and extreme stellar masses. We worked with two stellar mass functions, from \cite{shuntov25, Navarro-Carrera24}, and obtained estimates of the  extreme values of the $M_{\bh}/M_*$ ratio. We presented two pipelines. 

In Pipeline I, based on Gaussian convolution, the extreme values of $M_{\bh}$ were derived from those of $M_*$ via  log-linear correlations presented in \cite{Li25, pacucci23}. The resulting extreme ratios from these distributions are shown in \fref{fig:pip1_evs} and \fref{fig:pip1_ratio}.

In particular, the correlation from \cite{pacucci23} gave a low median ratio of $\approx 0.02$  with a large scatter. The data from \cite{kokorev, yue, matthee} are in agreement with the EVS prediction at $95^{\rm th}$ percentile, which extends to ratio as large as $\sim0.4$.



In the Pipeline II, the ratio of extremes, $\mu=M_{\text{BH,max}}/M_{*,\text{max}}$, was presented as a diagnostic measure for the tails of the mass functions. Samples were drawn from separate extreme-value distribution for the numerator and denominator of $\mu$. The black holes mass function used was that of \cite{Taylor2024}, which remains a subject of some debate. Analogous to Pipeline I, the main results for this Pipeline are shown in \fref{fig:pip2_evs} and \fref{fig:pip2_ratio}.

When averaging over the redshift range 4 - 8, we found $\mu\approx0.1$  and $0.24$ assuming the \cite{shuntov25} and \cite{Navarro-Carrera24} mass function respectively. Whilst Pipeline II generally gives a larger ratio and smaller scatter than Pipeline I, there are some discrepancies with the ratios observed in \cite{yue}.



The main results of our work are summarised in \tref{tab:summary}.

\subsection{Discussion}

\noindent\textit{Redshift evolution of the ratio:} In Pipeline I, we found little evolution for the extreme ratio $M_{\bh}/M_*$, although any evidence may be masked by the large scatter. Pipeline II gives a stronger evolution but this may be driven by the high mass end of the Taylor mass function.  \cite{roberts26} found little redshift evolution for the ratio based on the scaling in \cite{Li25,reines} for $z\lesssim5.5$. The  \cite{pacucci23} relation was shown to exhibit stronger redshift evolution. It is the median values of the coefficients at $z \sim 5$ which we used in \fref{fig:pip1_ratio} to compare with the data from \cite{yue}.

 Recently, \cite{wu26} examined the ratio $M_{\bh}/M_*$ over $ 1 < z < 10$ in a suite of simulations, finding a monotonic increase towards a broad peak at $\sim0.15$ at $z=7-10$. At $z = 5$, they found $M_{\bh}/M_* \approx 0.07$, which we find to be between the extreme values in our two Pipelines.  

\noindent\textit{Extreme ratios at $z>8$:} A quasar at $z \sim 10$ with $M_{\bh}/M_*$ as high as $M_{\bh}/M_* \sim 1$ was reported in \cite{bogdan}. If confirmed, this 
exceeds the expected value of the ratio derived from the EVS formalism which only relies on the empirically fitted Schechter mass functions. A range of additional factors could modify these mass functions to drive up the EVS-derived ratio at higher redshifts, including a heavy black hole seed, bursts of super-Eddington accretion, and suppression of star formation with AGN feedback and duty cycle.


\noindent\textit{Complex AGN physics:} One  of the limitations of the present analysis is that the BH mass function alone may not uniquely constrain the underlying black-hole growth history. Different combinations of accretion-rate distributions and AGN duty cycles can, in principle, produce similar SMBH mass functions, leading to a degree of degeneracy in the inferred growth scenarios \citep{Shankar_ea2010, Delvecchio_ea2020}. Additional observables, particularly measurements of AGN clustering, can provide complementary information on the halo occupation and active fraction of SMBHs, thereby helping to break these degeneracies. Unfortunately, robust clustering measurements of AGNs at the redshifts considered in this work remain limited. Future high-redshift surveys will therefore offer an important opportunity to jointly constrain both the growth history and the duty cycle of SMBHs.

The probabilities derived in this work should therefore be interpreted within the framework of independent-block EVS. While clustering and duty-cycle uncertainties may affect the absolute probabilities of the most massive BHs, and hence the ratio $M_{\bh}/M_*$, they are expected to influence all growth scenarios in a broadly similar manner. Consequently, the broad conclusions of this work, which are based on the relative comparison between different BH growth models, are unlikely to be significantly  altered by clustering.

\noindent\textit{Future work:} 
Our EVS framework could be used to calculate $M_{\bh}/M_*$ at lower redshifts ($z\sim1$), where black holes and star formation couple more strongly through feedback-regulated co-evolution {\citep{Silk_ea2024}}. However, this is limited by systematic uncertainties in black-hole mass measurements. Applying our formalism to local and high-redshift galaxies would show how extreme-value statistics can trace regimes in which black hole growth and stellar assembly decouple, providing key insight into the physics of early galaxy formation.







\section*{Acknowledgements}
We thank the referee for their valuable suggestions. Cameron Heather is supported by the Warwick Mathematics Institute Centre for Doctoral Training, and gratefully acknowledges funding from the University of Warwick and the UK Engineering and Physical Sciences Research Council (grant number: EP/W524645/1). This work was supported by Naresuan University (NU), and National Science, Research and Innovation Fund (NSRF). Grant NO.R2569B119.

\section*{Data Availability}
The data underlying Figs. 5–8 and Tables 1–2 of this article are available at Zenodo  DOI:\dataset[10.5281/zenodo.21842468]{https://10.5281/zenodo.21842468}. Additional data used in this work are available on reasonable request to the corresponding author.

\bibliography{references}{}
\bibliographystyle{aasjournalv7}

\appendix
\section{Extreme ratio with Li et al. correlation}\label{appendix}
\begin{figure}[htb!]
    \centering
    \includegraphics[width = \linewidth]{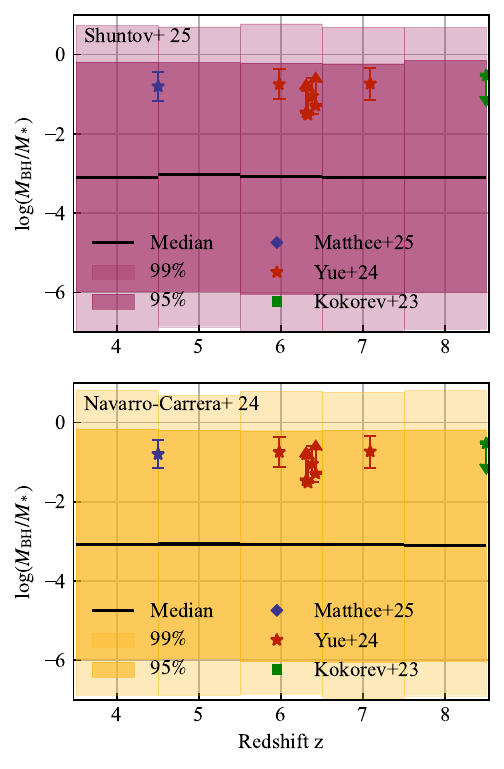}
    \caption{The EVS prediction for the extreme ratio $M_{\bh}/M_*$ for the redshift range $z = 3.5 - 8.5$ using Pipeline I and the \cite{Li25} scaling relation. The stellar mass functions used are  \cite{shuntov25} (top) and \cite{Navarro-Carrera24} (bottom). The solid black line corresponds to the median, whilst the darker/lighter shaded area corresponds to the $95^{th}$/$99^{th}$ percentile. The data points are from \cite{matthee, yue, kokorev}.
    }\label{fig:li_evs}
\end{figure}

Here we include the results from Pipeline I in \sref{sec:gauss_evs} using the log-linear correlation from \cite{Li25}. 
The EVS prediction for ratio $M_{\bh}/M_*$ is shown in \fref{fig:li_evs}. While the data points are within the $95\%$ confidence interval for \cite{Li25}, the range is far too large to be used meaningfully in predicting the extreme ratio. This is likely due to the various biases in the selection effects, causing the scatter in the scaling relation to increase.

\end{document}